\documentclass{article}
\usepackage{graphicx}
\usepackage{amssymb}

\begin{document}

\title{Modeling the response of a recovering SiPM}

\date{\today}

\author{Daniel Jeans,\\
Department of Physics, \\
Graduate School of Science, \\
The University of Tokyo, \\
7-3-1 Hongo,
Bunkyo-ku, \\
Tokyo 113-0033,
Japan. \\
jeans@icepp.s.u-tokyo.ac.jp
}

\maketitle

\abstract{
We develop from first principles a model to describe the average response of SiPM devices
which takes into account the recovery of pixels during the incoming light pulse.
Such effects can significantly affect SiPM response when exposed to a large number of photons.
}

%

\section{Introduction}

Silicon Photomultipliers (SiPM) 
have revolutionised photon detection in several fields including particle physics and medical imaging,
thanks to their minute size, good photon detection efficiency, insensitivity to magnetic fields,
low operation voltage and modest cost.

In this paper we present a model of the average SiPM response which is particularly relevant
when the length of the input light pulse is similar or longer in length than the
SiPM pixel recovery time.
The recovery times for real SiPM devices can be as short as a few ns~\cite{Uozumi, gruber2},
less than the emission time of many of the scintillators and wavelength shifters 
to which SiPMs are commonly coupled.
The effect of pixel recovery 
becomes important
when the probability for a pixel to be hit
by more than a single photon becomes significant, i.e. at high input light levels.
Such high light levels are, for example, encountered in applications to electromagnetic
calorimetry, in which SiPMs must operate over a large dynamic range, capable of measuring
signals produced by both minimum ionising particles and the cores of electromagnetic showers,
and may operate near saturation in the latter case.

The response of SiPMs under such high light conditions deviates significantly
from the expectations for a purely digital device without pixel recovery~\cite{gruber1}.
A semi-empirical model describing these effects has recently been developed~\cite{kotera}.
In this paper, we develop a model from first principles in which an exponential light pulse 
(as expected from a simple model of light emission from a scintillator or wavelength shifter) 
is input to the SiPM,
and in which SiPM pixels recover exponentially after an avalanche. 
Our model does not consider other SiPM characteristics such as after-pulses and pixel cross-talk,
treating only the effects of pixel recovery.
We formulate the model in section~\ref{sec:model}, compare it to Monte Carlo simulation in section~\ref{sec:toy},
and conclude in section~\ref{sec:conclude}.

\section{Model}
\label{sec:model}

\subsection{Balls and Bins - Photons and Pixels}
\label{sec:balls}

It is well-known (e.g. \cite{balls}) that if $m$ balls are randomly distributed among $N$ bins, 
then the expected number of bins $p$ containing exactly $k$ balls is given by
\begin{eqnarray}
%
%
p ( N, m, k ) & = & N  \ \frac{m!}{ (m-k)! k!} \ \Big[ 1 - \frac{1}{N} \Big] ^ {m-k} \ \Big[ \frac{1}{N} \Big] ^ k \\
%
\label{eqn:n}
& = & \frac{N}{k!} \ \Big[ \frac{m}{N} \Big]^k \ e^{-m/N} .
\end{eqnarray}
We can directly apply these results to populating $N$ SiPM pixels with $m$ photons.
The average number of SiPM pixels hit by at least one photon when illuminated by $m$ photons 
\begin{equation}
\label{eqn:binary}
p(N,m,k>0) = N - p( N, m, 0 ) = N ( 1 - e^{-m/N} ),
\end{equation}
reproducing the well-known response of a SiPM with purely digital characteristics.

\subsection{Pixel recovery}

The charge produced by the avalanche induced within a SiPM pixel depends on the extent to which it is fully charged, and 
therefore on the time since it was last discharged.
We assume that the recovery of a pixel after firing is exponential, 
and that the released charge $Q$ depends on the time since last firing $\delta t$ as
\begin{equation}
Q = Q_0 \big[ 1 - e^{-\delta t / \tau_R} \big] ,
\label{eqn:dt}
\end{equation}
where $Q_0$ is the charge produced by a fully-charged pixel, and $\tau_R$ is the {\em recovery time} of the pixel, which
can, in a simple model, be estimated in terms of the quenching resistance $R$ and capacitance $C$ of the pixel: $\tau_R = RC$.

\subsection{Single pixel multi-photon response}
\label{sec:multiphoton}

In this section we derive the average charge produced when $k$ photons from an exponentially-distributed light pulse ($dn/dt \propto e^{-t/\tau_S}$)
fall on a single SiPM pixel.
The time one has to wait for the first of these $k$ photons to hit the pixel is exponentially distributed, with decay constant $\tau_S / k$; 
once this first photon has struck, the time before the second photon arrives is also exponentially distributed, but with time constant $\tau_S / (k-1)$.
In general the time between the $(j-1)^{th}$ and $j^{th}$ photons' arrivals
is characterised by a decay constant $\tau_j = \tau_S / (k-j+1)$.

The first photon will always release a charge $Q_1 = Q_0$, while
the average charge released by the $j^{th}$ photon ($j \geq 2$) can be written as
\begin{eqnarray}
Q_j & = &  Q_0 \ \int_{0}^{\infty} { \frac{1}{\tau_j} \ e^{-t/\tau_j} \ ( 1 - e^{-t/\tau_R} ) } \ dt \\
    & = &  Q_0 \ \Big[ 1 - \frac{\tau_R}{\tau_R + \tau_j} \Big] \\
    & = &  Q_0 \ \Big[ 1 - \frac{\zeta}{\zeta + (k-j+1)^{-1}} \Big]
\end{eqnarray}
where $\zeta \equiv \tau_R / \tau_S$.
The average total charge produced by $k$ photons (when $k \geq 1$) in a single pixel is then
\begin{eqnarray}
 Q^{\rm pix}_{\rm exp}  & = & Q_1 + \sum_{j=2}^{k} Q_j \\
\label{eqn:Q}
              & = & Q_0 \bigg( 1 + \sum_{j=2}^{k}  \Big[ 1 -  \frac{\zeta}{\zeta + (k-j+1)^{-1}} \Big]  \bigg) 
\end{eqnarray}
%
%
where the sum only enters if $k \geq 2$.
The sum can be written in terms of the digamma function $\digamma ( x )$, leading to
\begin{equation}
 Q^{\rm pix}_{\rm exp} = Q_0 \bigg( 1 + \frac{1}{\zeta}  \Big[ \digamma(1 - k - 1/\zeta) - \digamma(-1/\zeta) \Big] \bigg) .
\footnote{Since $\digamma(x)$ is ill-defined for x=0, -1, -2, -3..., this form cannot be calculated for exactly integer values of $1/\zeta$. 
Applying a small offset to $\zeta$ in such cases restores good behaviour.
The difference of the two digamma functions remains well-defined.
}
\end{equation}


If the input light pulse has a different distribution, or the pixel recovers differently to the simple model considered here,
the above analysis can be adapted accordingly. As an example, if the $k$ input photons are distributed uniformly over a time $T$
(in which case the average time between photon arrivals is always $T/(k+1)$) and the pixel recovery follows $Q = Q_0 [1 - f(\delta t)]$,
then the average total pixel charge
\begin{equation}
 Q^{\rm pix}_{\rm uniform} = Q_0 \bigg( 1 +  (k-1) \Big[ 1 - f \big( T/(k+1) \big) \Big] \bigg).
\end{equation}

\subsection{Total SiPM response}

We then combine equations \ref{eqn:n} and \ref{eqn:Q}
to derive the total response of a SiPM with
$N$ pixels hit by $m$ photons from an exponential light pulse:
\begin{equation}
\label{eqn:final}
Q^{\rm SiPM} (N, m, \zeta) = \sum_{k=1}^{m} \Big[ p (N, m, k) \cdot Q^{\rm pix}_{\rm exp} ( k, \zeta ) \Big] . 
\end{equation}
The function $Q^{\rm SiPM}$ depends only on the total number of SiPM pixels $N$, the number of incident photons $m$, 
and $\zeta$, the ratio of the SiPM pixel recovery time to the decay time of the input light pulse.

\section{Comparison to Monte Carlo simulation}
\label{sec:toy}
\begin{figure}
\center
\includegraphics[width=0.46\textwidth]{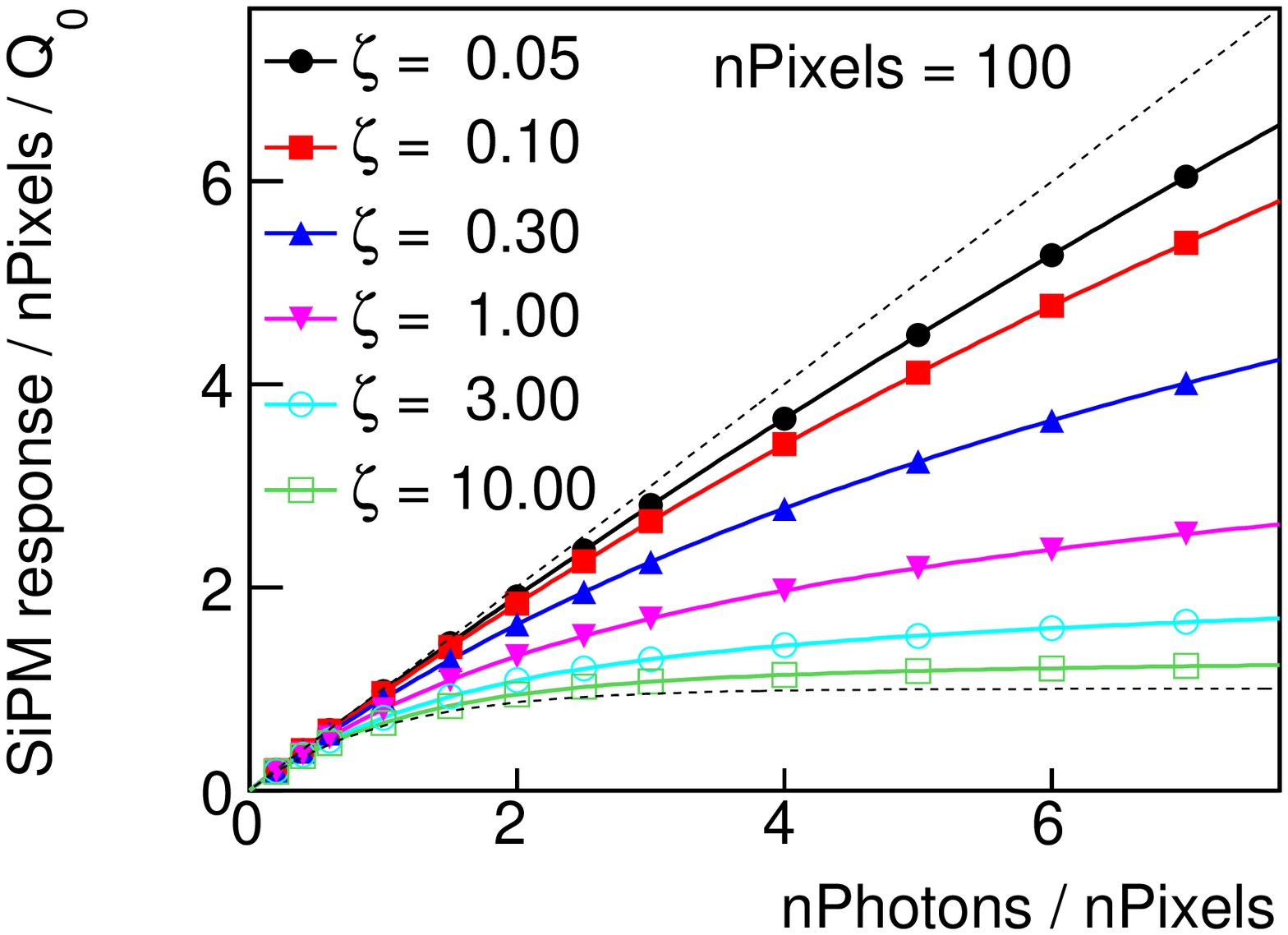}
\includegraphics[width=0.46\textwidth]{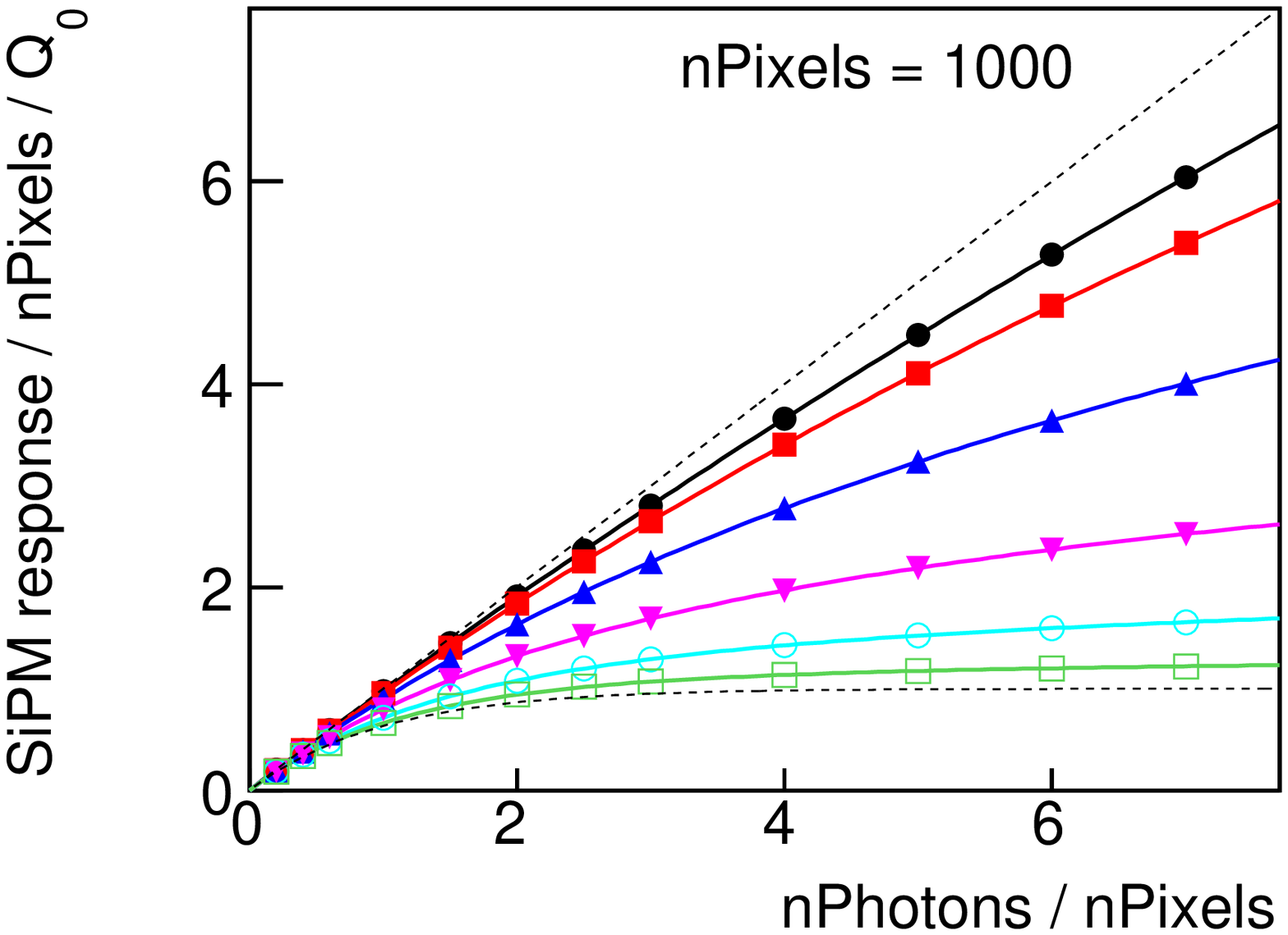} \\
\includegraphics[width=0.46\textwidth]{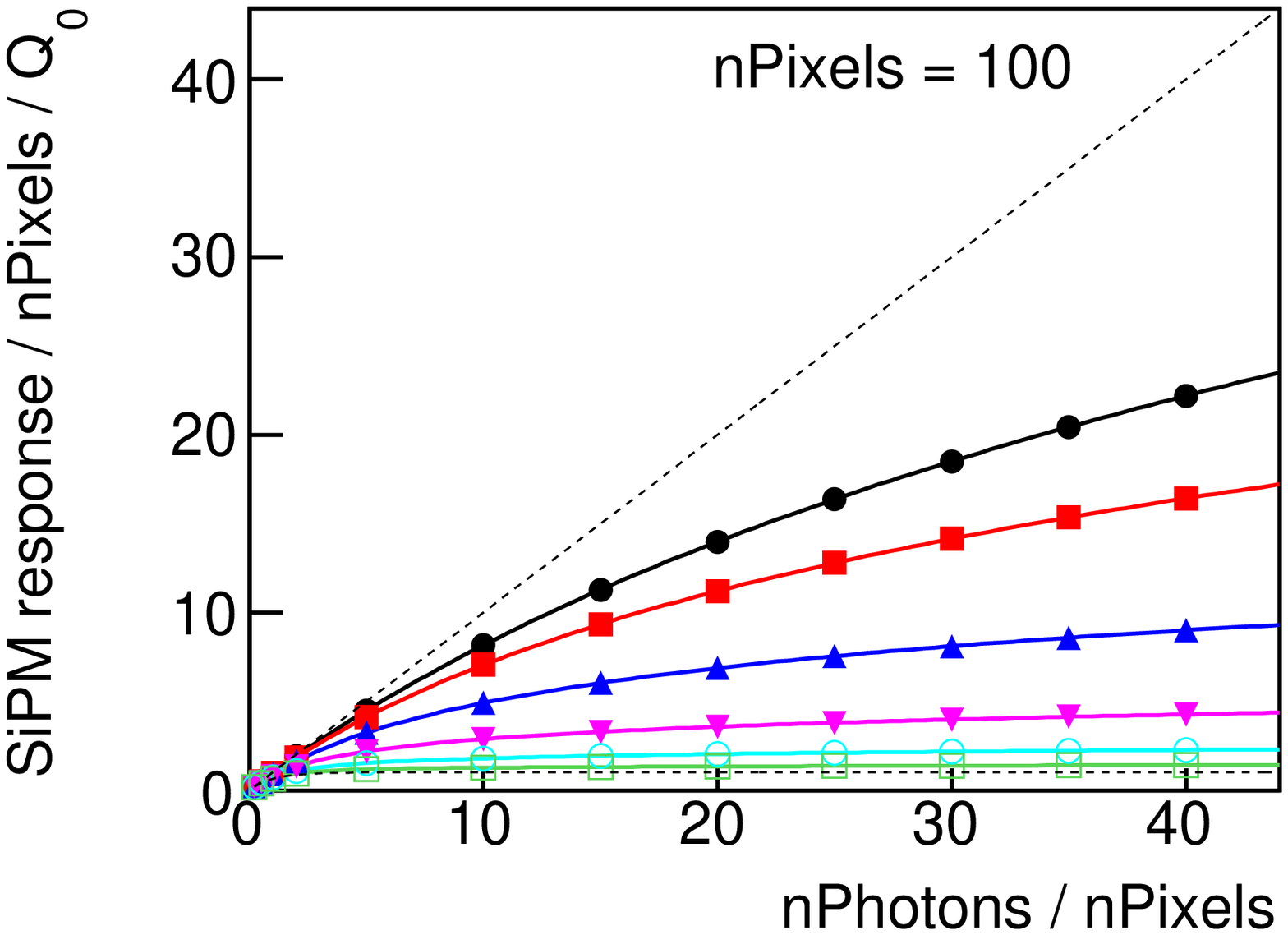}
\includegraphics[width=0.46\textwidth]{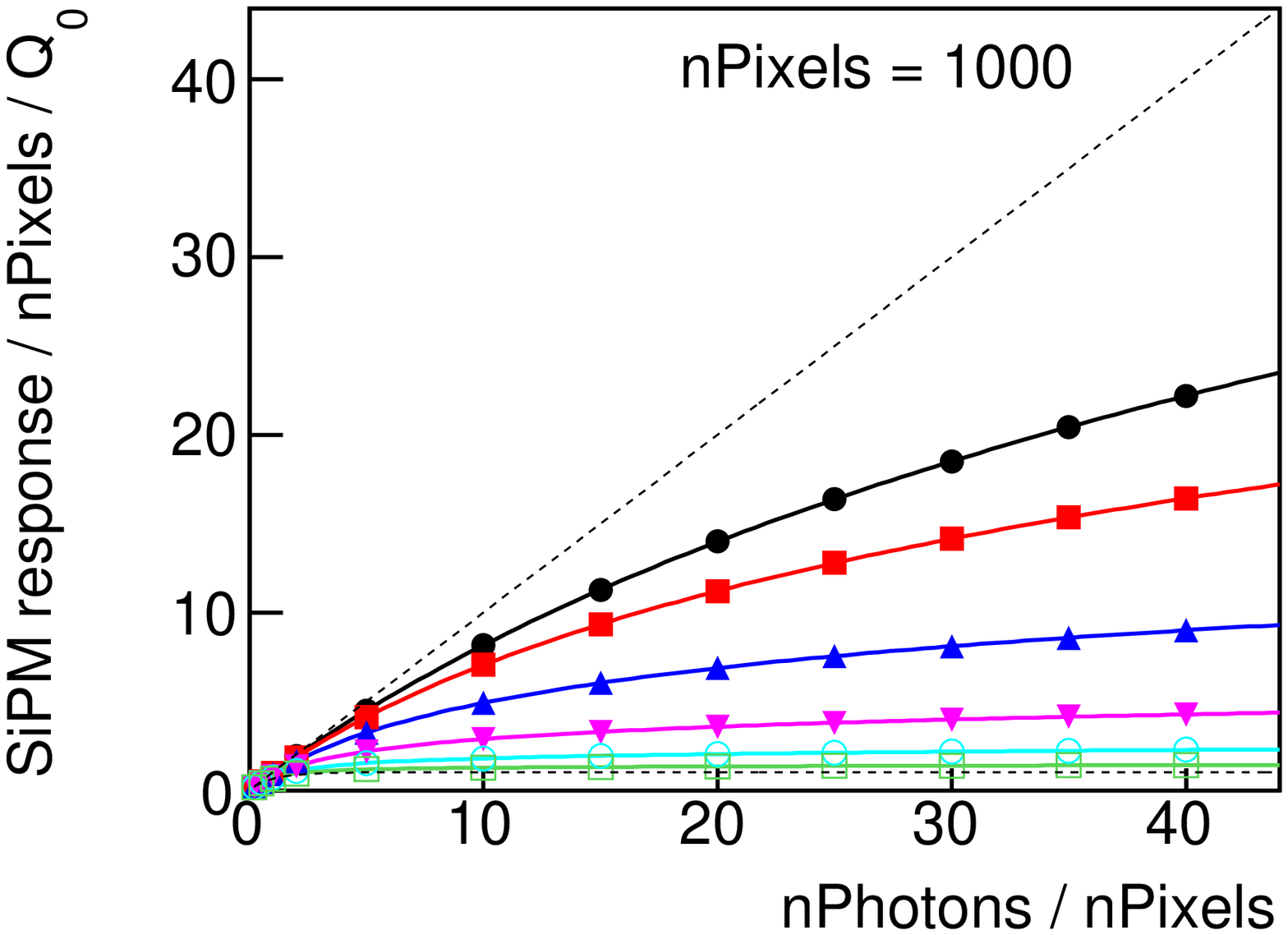} \\
\includegraphics[width=0.46\textwidth]{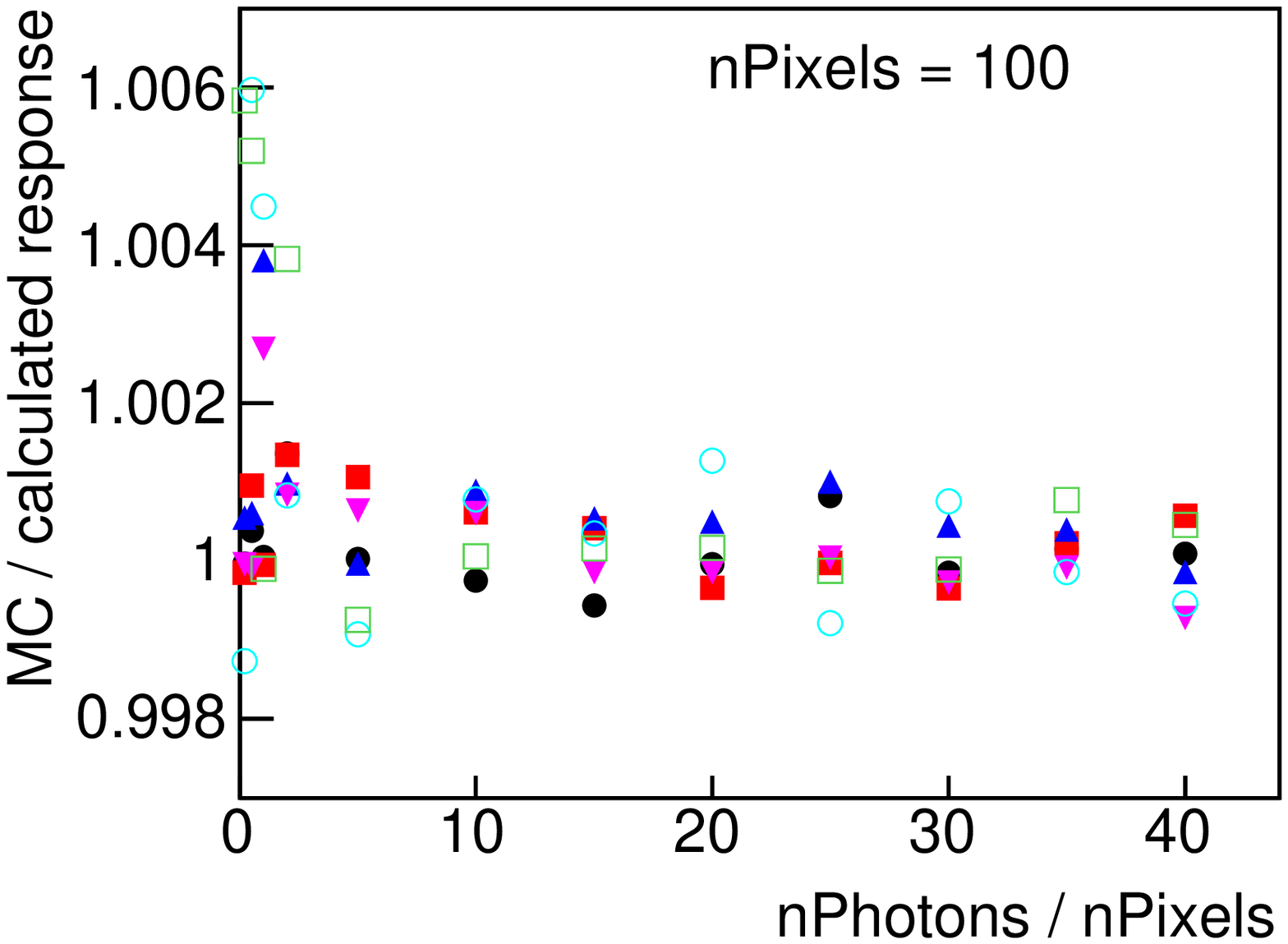} 
\includegraphics[width=0.46\textwidth]{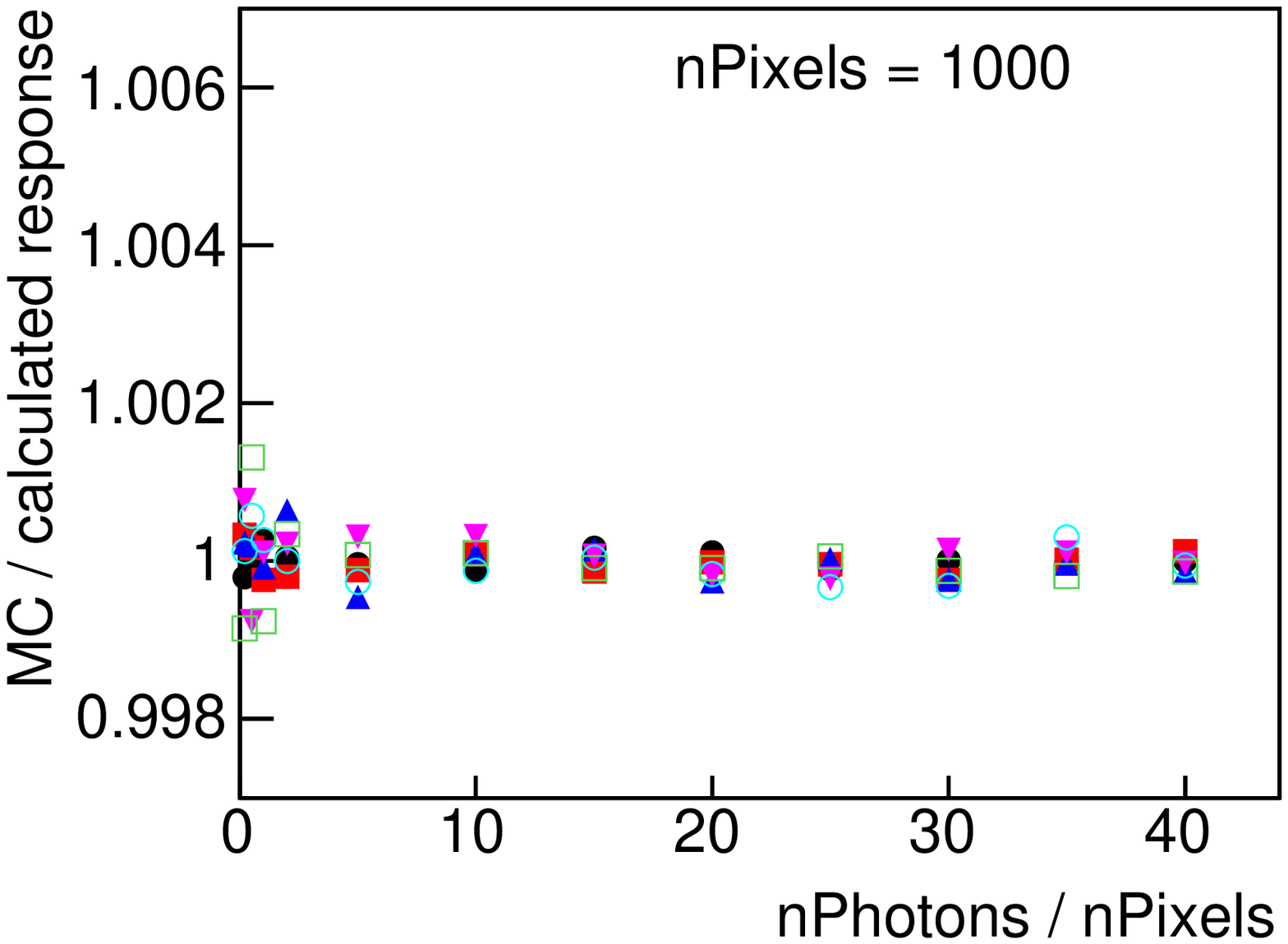} 
\caption{
Response of a 100 (left) and 1000 (right) pixel SiPM when illuminated by an exponentially distributed light pulse.
The upper (middle) plots show the response in a limited (wide) range of input photons at different values of $\zeta$,
as estimated by simulation (points) and calculated using equation~\protect\ref{eqn:final} (continuous lines).
The dashed lines show the two extreme cases of a perfectly linear response and of a perfectly digital SiPM (
described by 
equation~\protect\ref{eqn:binary}).
The lower plots show the ratio of the simulated to the calculated response.
}
\label{fig:toy}
\end{figure}

The above calculation uses the {\em expected} number of pixels containing a certain number of photons
and the {\em average} charge they release to arrive at the final result. 
In this section we compare the calculation to a full Monte Carlo simulation in which these averages are not performed.
The simulation  randomly distributes a given number of photons among a fixed number of pixels, 
also assigning to each photon a random time extracted from an exponential distribution.
The charge associated to each photon is assigned according to equation~\ref{eqn:dt}, taking
the time from the previous hit within the same pixel into account.
We have simulated 100 and 1000 pixel SiPMs with different $\zeta$ factors over a wide range of input
photon number. The results are presented in figure~\ref{fig:toy}, which also shows the 
corresponding functions defined by equation~\ref{eqn:final}.
The prediction of equation~\ref{eqn:final} agrees with the results of the Monte Carlo model to within a few 
per mille. 


\section{Conclusion}
\label{sec:conclude}

We have developed a model for the response of a SiPM when exposed to a light pulse extended in time,
starting from the statistical properties of assigning photons to pixels, the average
distribution in time of photons within a light pulse, and a simple exponential model of SiPM pixel recovery.
In the case of an exponential input light pulse, the resulting response equation depends only on the
total number of SiPM pixels and input photons, and the ratio of the input light pulse decay constant and
the recovery time of the SiPM. Adjusting this ratio varies the response from that of a purely
digital SiPM (with no pixel recovery) to a perfectly linear one (with immediate pixel recovery).


The model describes the general features observed in real SiPM devices, particularly at high
photon exposure (as shown for example in~\cite{gruber1, kotera}) where the response significantly
exceeds the maximum response expected from a simple digital SiPM response model.
We expect that to accurately describe SiPM response, the present model should be extended to include
descriptions of inter-pixel cross-talk and after-pulsing. Effects of the geometrical acceptance and
quantum efficiency of the SiPM can in principle be taken account of by appropriately scaling the number of input
photons of the present model.

\section*{Acknowledgements}
I thank K.~Kotera and S.~Uozumi for useful discussions.
This work was funded by the MEXT KAKENHI Grant-in-Aid for Scientific Research on Innovative Areas, no. 23104007.

\appendix
\section{Example code}
The following code implements the model of equation~\ref{eqn:final}.

\small
\begin{verbatim}
double factorialFunction(int i) { 
  double fac(1);
  while (i > 1) fac*=i--;
  return fac;
}
double expectedNumberOfPixels(int nPixels, int nPhotons, int nPixelPhotons) {
  double t(nPhotons); t/=nPixels;
  return nPixels*pow(t,nPixelPhotons)*exp(-t)/factorialFunction(nPixelPhotons);
}
double singlePixelCharge(int nPhotons, double zeta) {
  double charge(0);
  if (nPhotons>0) charge+=1;
  if (nPhotons>1) {
    for (int j=2; j<=nPhotons; j++) 
      charge+=1-zeta/(zeta+pow(nPhotons-j+1,-1)) ;
  }
  return charge;
}
double mppcResponse(int nPixels, int nPhotons, double zeta) {
  double totCharge(0);
  double totPixels(0);
  for (int i=1; i<nPhotons; i++) {
    double expPix=expectedNumberOfPixels(nPixels, nPhotons, i);
    totPixels+=expPix;
    if (expPix>1e-10) totCharge+=expPix*singlePixelCharge(i, zeta);
    if (nPixels-totPixels<1e-5) break;
  }
  return totCharge;
}

\end{verbatim}

\end{document}